\pdfoutput=1

\newcommand*{\ATLASLATEXPATH}{latex/}
\documentclass[cernpreprint, texlive=2016, coverpage=false, UKenglish]{\ATLASLATEXPATH atlasdoc}

\usepackage{\ATLASLATEXPATH atlaspackage}
\usepackage{\ATLASLATEXPATH atlasbiblatex}
\usepackage{\ATLASLATEXPATH atlasphysics}

\graphicspath{{logos/}{figures/}}

\usepackage{multirow}
\usepackage{makecell}

\AtlasTitle{Search for Higgs boson pair production in the $WW^{(*)}WW^{(*)}$ decay channel using ATLAS data recorded at $\sqrt{s}=13$~TeV}

\AtlasJournalRef{JHEP 05 (2019) 124}
\AtlasDOI{10.1007/JHEP05(2019)124}

\AtlasAbstract{%
  A search for a pair of neutral, scalar bosons with each decaying into two $W$ bosons is presented using 
  36.1 fb$^{-1}$ of proton--proton collision data at a centre-of-mass energy of 13~TeV recorded with the 
  ATLAS detector at the Large Hadron Collider.  This search uses three production models: non-resonant 
  and resonant Higgs boson pair production and resonant production of a pair of heavy scalar particles.  
  Three final states, classified by the number of leptons, are analysed: two same-sign leptons, three 
  leptons, and four leptons. No significant excess over the expected Standard Model backgrounds is 
  observed.  An observed (expected) 95\% confidence-level upper limit of 160~(120) times the Standard 
  Model prediction of non-resonant Higgs boson pair production cross-section is set from a combined analysis 
  of the three final states.  Upper limits are set on the production cross-section times branching ratio 
  of a heavy scalar $X$ decaying into a Higgs boson pair in the mass range of 260~GeV~$\le m_X \le$~500~GeV 
  and the observed (expected) limits range from 9.3~(10)~pb to 2.8~(2.6)~pb.  Upper limits are set on the 
  production cross-section times branching ratio of a heavy scalar $X$ decaying into a pair of heavy 
  scalars $S$ for mass ranges of 280~GeV~$\le m_X \le$~340~GeV and 135~GeV~$\le m_S \le$~165~GeV 
  and the observed (expected) limits range from 2.5~(2.5)~pb to 0.16~(0.17)~pb.
}

\author{The ATLAS Collaboration}
\AtlasRefCode{HIGG-2016-24}
\PreprintIdNumber{CERN-EP-2018-227}
\AtlasDate{\today}
\AtlasJournal{JHEP}

\hypersetup{pdftitle={ATLAS document},pdfauthor={The ATLAS Collaboration}}

\begin{document}

\maketitle

\tableofcontents

\section{Introduction}
\label{sec:Introduction}

A scalar boson was discovered by the ATLAS and CMS collaborations~\cite{HIGG-2012-27,CMS-HIG-12-028} 
in 2012. It has been shown to have properties consistent with those predicted for the Standard Model (SM) 
Higgs boson, $H$, through spin and coupling 
measurements~\cite{HIGG-2015-07,HIGG-2013-17,CMS-HIG-14-018,HIGG-2015-06,HIGG-2015-07,HIGG-2018-13,CMS-HIG-17-035,HIGG-2018-04,CMS-HIG-18-016}.  
These measurements are based on production of the Higgs boson via gluon--gluon fusion, vector-boson 
fusion and in association with a $W$ or $Z$ boson or a top quark pair. The SM predicts non-resonant Higgs boson 
pair production via top quark loops as well as through self-coupling. The SM $HH$ production cross-section is 
computed to be 33.4~fb~\cite{eftreweight,HiggsXSec} at next-to-next-to-leading order (NNLO) in QCD, including 
resummation of soft-gluon emission at next-to-next-to-leading-logarithmic (NNLL) accuracy for $m_{H} = 125.09$~GeV. 
The actual production rate could be larger than that predicted in the SM due to a variety of Beyond the Standard 
Model (BSM) physics effects. One such extension includes a modification to the SM Higgs self-coupling, 
$\lambda_{HHH}$, and another the existence of a new heavy resonance which decays into a pair of Higgs bosons.  
An important Higgs boson decay channel is $H \rightarrow VV^{(*)}$ in which $V$ can be either a $W$ or $Z$ boson, 
on or off-shell, and this paper focuses on the $4W$ final state~\cite{4WTheory} in both SM and BSM $HH$ production 
scenarios. 

This work investigates $HH$ production through three different processes. The first is (\ref{eq:nonresonant}) the 
SM $HH$ production (non-resonant $HH$). The second and third are both BSM processes inspired by an extended 
Higgs sector, such as a two-Higgs-doublet model~\cite{2HDMTheory}, in which a neutral heavy Higgs boson, 
$X$~\cite{ExoticHiggsTheory} is produced  and decays either (\ref{eq:resonant}) directly into two SM Higgs bosons 
(resonant $HH$) or (\ref{eq:XS}) into a pair of new scalar bosons, $S$ ($X\rightarrow SS$), each of which in turn 
decays to other SM particles with the same mass-dependent branching ratios of the SM $H$. The reactions 
considered in this work are: 

\begin{align}
pp &\rightarrow HH \rightarrow WW^{(*)}WW^{(*)} \text{ (non-resonant, SM),} \label{eq:nonresonant}\\
pp &\rightarrow X \rightarrow HH \rightarrow WW^{(*)}WW^{(*)} \text{ (resonant, BSM), and } \label{eq:resonant}\\
pp &\rightarrow X \rightarrow SS \rightarrow WW^{(*)}WW^{(*)} \text{ ($X\rightarrow SS$, BSM).} \label{eq:XS}
\end{align}

The measured final states encompass multiple combinations of leptons and hadrons:

\begin{align}
WW^{(*)}WW^{(*)} &\rightarrow \ell \nu + \ell \nu  + 4q, \nonumber \\
WW^{(*)}WW^{(*)} &\rightarrow \ell \nu  + \ell \nu  +  \ell \nu  + 2q, \text{ or} \nonumber\\
WW^{(*)}WW^{(*)} &\rightarrow \ell \nu +  \ell \nu  +  \ell \nu   + \ell \nu \nonumber
\end{align}

where $\ell$ is either an electron or a muon, $q$ refers to quark and anti-quark decay products from the 
hadronically decaying $W$ boson(s), and $\nu$ represents a neutrino, which results in missing transverse 
momentum. Therefore, three final states are searched for with two, three, or four leptons (plus missing energy 
and multiple jets), which allow any of the mentioned production modes to be probed.

The production of a new $X$ scalar (\ref{eq:resonant}) would be seen as a local excess in the reconstructed 
di-Higgs mass spectrum.  It is assumed in this work that $m_X > 2m_H$ such that both $H$ are produced 
on their mass shell.  In the other extended Higgs sector model (\ref{eq:XS}) $X\rightarrow SS$ is assumed 
to be the dominant $X$ decay mode. In this scenario, the $WW^{(*)}WW^{(*)}$ channel is the dominant decay 
mode for the mass ranges 270~GeV~$< m_X < 2m_t$ and 135~GeV~$< m_S < m_X/2$, where $m_t$, 
$m_X$ and $m_S$ are the mass of the top quark, $X$, and $S$ scalars, respectively.  The mass range 
$m_X > 2m_t$, where $X\rightarrow t\bar{t}$ is expected to dominate, is not considered. It is assumed that 
$m_S > 135$~GeV such that $S\rightarrow WW^{(*)}$ is the dominant decay mode. It is also assumed that 
$m_S < m_X/2$ such that both $S$ bosons are produced on their mass shell.

Previous searches were performed for resonant and non-resonant $HH$ production using various channels, 
such as $bb\gamma\gamma$~\cite{HIGG-2016-15,CMS-HIG-17-008}, $bbbb$~\cite{EXOT-2016-31,CMS-HIG-17-009,CMS-HIG-17-017}, 
$bbVV$~\cite{CMS-HIG-17-006}, $bb\tau\tau$~\cite{HIGG-2016-16,CMS-HIG-17-002} and 
$WW\gamma\gamma$~\cite{HIGG-2016-20}, with data from the ATLAS and CMS experiments.  Additionally, 
a combination of channels has been performed using data from the CMS experiment~\cite{cms-hhcomb}.  
This paper describes a search for resonant and non-resonant Higgs boson pair production in the 
$HH\rightarrow WW^*WW^*$ decay channel and for an extended Higgs sector with the decay of 
$X\rightarrow SS\rightarrow WW^{(*)}WW^{(*)}$. The analysis is divided into three independent channels 
depending on the number of light leptons ($e$ or $\mu$) from leptonic decays of $W$ bosons, and then 
statistically combined to give the final result.

This paper is organised as follows. Data and simulation samples are described in Section~\ref{sec:DataMC}.  
The object reconstruction and selection are outlined in Section~\ref{sec:Objects}. Section~\ref{sec:EventSelection}  
details the event selection for each of the three final states analysed. The background estimation and the 
systematic uncertainties are described in Section~\ref{sec:BackgroundEstimation} and Section~\ref{sec:Systematics}, 
respectively. The results of this analysis are presented in Section~\ref{sec:Results} and summarised in 
Section~\ref{sec:Conclusion}. Finally, the Appendix lists the lepton pairing strategy used in each channel, the final 
event selection criteria and the corresponding acceptance and selection efficiencies.

\section{Data and simulation samples} 
\label{sec:DataMC}

The data were collected with the ATLAS detector in 2015 and 2016 using $pp$ collisions produced 
at $\sqrt{s}=13$~TeV at the Large Hadron Collider (LHC), corresponding to an integrated luminosity 
of 36.1~fb$^{-1}$~\cite{DAPR-2013-01}.  The ATLAS detector is described in detail in Ref.~\cite{PERF-2007-01}. 
Only data-taking periods in which all relevant detector systems are operational are used.

Samples simulated using Monte Carlo (MC) techniques are used to estimate the signal acceptance 
and selection efficiency. Simulated samples are also used to estimate the acceptance and selection 
efficiency for various background processes which contribute prompt leptons from $W$ or $Z$ boson 
decay and leptons originating from photon conversion.  Backgrounds due to electrons with misidentified 
charge and jets misidentified as leptons are estimated using data-driven techniques, as described in 
Section~\ref{sec:BackgroundEstimation}. 

The non-resonant $gg\rightarrow HH$ and resonant $gg\rightarrow X \rightarrow HH$ signal samples 
in which $H$ is constrained to decay into $WW^*$ are generated using 
\MGMCatNLO~\cite{madgraph5amcnlo,syscalc} with the CT10 parton distribution function (PDF) 
set~\cite{PDF:CT10} and the parton shower is modelled by \Herwigpp~\cite{herwigpp} with the UEEE5 
set of tuned parameters (tune) for the underlying event~\cite{herwigtune} and the CTEQ6L1 PDF 
set~\cite{Pumplin:2002vw}.  In resonant production, $X$ decays into a pair of SM Higgs bosons with 
a negligible width compared to the experimental mass resolution.  Various resonance mass hypotheses, 
$m_X$, are considered: 260, 300, 400, and 500~GeV. The branching ratio $\mathcal{B}(X\rightarrow HH)$ 
is assumed to be one.  Samples of $X\rightarrow SS\rightarrow WW^{(*)}WW^{(*)}$ events produced by 
gluon--gluon fusion are generated at leading order (LO) using \PYTHIAV{8} with the NNPDF2.3LO PDF 
set~\cite{PDF:NNPDF23} such that both the $X$ and $S$ scalars are assumed to have narrow decay 
widths. The mass hypotheses are selected to scan a range of both $m_X$ and $m_S$. In the first scan, 
$m_S$ is fixed to 135~GeV for samples with $m_X$~=~280, 300, 320, and 340~GeV.  In the second scan, 
$m_X$ is fixed to 340~GeV for samples with $m_S$~=~135, 145, 155, and 165~GeV. The branching ratio 
$\mathcal{B}(X\rightarrow SS)$ is assumed to be one and the branching ratio 
$\mathcal{B}(S\rightarrow WW^{(*)})$ is assumed to be the mass-dependent expected branching ratios of 
the SM Higgs boson.

Multi-boson ($VV/VVV$) and $V\gamma$ background samples are generated at next-to-leading-order 
(NLO) using \SHERPAV{2.1}~\cite{sherpa}. The $V$+jets samples are generated at NLO with \SHERPAV{2.2}. 
The CT10 PDF set is used for these samples.  The $VH$ background sample is generated at leading-order 
(LO) using \PYTHIAV{8} with the NNPDF2.3LO PDF set.  The \ttbar background sample is generated at NLO 
using \POWHEGBOXV{2.0}~\cite{powhegbox} interfaced with \PYTHIAV{8} with the NNPDF2.3LO PDF set.  
Single-top background samples are generated at NLO using \POWHEGBOXV{2.0} interfaced with 
\PYTHIAV{6.4}~\cite{pythia6} with the CT10 PDF set.  The $\ttbar V$ background sample is generated at NLO 
using \MGMCatNLO interfaced with \PYTHIAV{8} with the NNPDF2.3LO PDF set. The $\ttbar H$ background 
sample is generated at NLO using \MGMCatNLO interfaced with \Herwigpp with the NNPDF3.0~\cite{PDF:NNPDF30} 
PDF set.  The simulated samples of \ttbar, $\ttbar H$, $\ttbar V$, and $VV$ are described in more detail in 
Refs.~\cite{ATL-PHYS-PUB-2016-004,ATL-PHYS-PUB-2016-005,ATL-PHYS-PUB-2016-002}.

The standard ATLAS detector simulation~\cite{SOFT-2010-01} based on \textsc{Geant4}~\cite{geant4} 
is used for background simulated samples. For signal events, the calorimeter simulation is replaced with 
the fast ATLAS calorimeter simulation~\cite{ATL-SOFT-PUB-2014-01} that uses a parameterised detector 
response.  Soft collisions generated using \PYTHIAV{8}~\cite{pythia8} with the CTEQ6L1 PDF set and the 
A2 tune~\cite{ATL-PHYS-PUB-2012-003} are overlaid on the hard-scatter processes. The number of in-time 
and out-of-time collisions per bunch crossing (pileup) is adjusted to that observed in data.

\section{Object selection}
\label{sec:Objects}

Electron candidates are reconstructed from energy clusters in the electromagnetic calorimeter that are associated 
with tracks reconstructed in the inner detector (ID). Electrons are identified using medium (tight) 
criteria~\cite{ATLAS-CONF-2016-024} for the four lepton channel (two and three lepton channels).  Electrons are 
required to have a transverse energy $\et>10$~GeV and be within the detector fiducial volume of $|\eta|<2.47$ 
excluding the transition region between the barrel and end-cap calorimeter, $1.37<|\eta|<1.52$.\footnote{ATLAS 
uses a right-handed coordinate system with its origin at the nominal interaction point (IP) in the centre of the detector 
and the $z$-axis along the beam pipe. The $x$-axis points from the IP to the centre of the LHC ring, and the $y$-axis 
points upwards. Cylindrical coordinates ($r$, $\phi$) are used in the transverse plane, $\phi$ being the azimuthal 
angle around the $z$-axis.  The pseudorapidity is defined in terms of the polar angle $\theta$ as 
$\eta =  - \ln \tan(\theta/2)$. Angular distance is measured in units of $\Delta R \equiv \sqrt{(\Delta\eta)^2 + (\Delta\phi)^2}$.}
Muon candidates are reconstructed by combining tracks reconstructed in the ID with tracks reconstructed 
in the muon spectrometer. Muons are identified using medium (tight) criteria~\cite{PERF-2015-10} for the 
four lepton channel (two and three lepton channels).  
Muons are required to have a transverse momentum $\pt>10$ GeV and $|\eta|<2.5$. Electrons are 
required to satisfy calorimeter and track isolation criteria and muons are required to satisfy a track 
isolation criterion.  The calorimeter (track) isolation requires that the total sum of cluster transverse 
energies (transverse momenta of tracks with $\pt>1$~GeV) in a surrounding cone of size $\Delta R=0.2$ 
around the lepton, excluding the cluster \et (track \pt) of the lepton from the sum, is less than 30\% (15\%) of 
the \pt of the lepton for the four lepton selection and 6\% for the two and three lepton selections.
 
Jets are reconstructed from calibrated topological clusters in the calorimeters~\cite{PERF-2014-07} using 
the \mbox{anti-$k_t$} algorithm~\cite{akt} with a radius parameter $R=0.4$. Jet energies are corrected for 
effects from the detector and from pileup~\cite{PERF-2016-04} using simulated and \textit{in situ} 
techniques~\cite{PERF-2016-04}. Jets are required to have $\pt$~$>$~25~GeV and $|\eta|$~$<$~2.5. Jets with 
$\pt<60$~GeV and $|\eta|<2.4$ are required to satisfy additional pileup rejection criteria~\cite{ATLAS-CONF-2014-018}. 
Jets containing $b$-hadrons are identified ($b$-tagged) using the MV2c10 multivariate discriminant~\cite{PERF-2016-05}. 
The $b$-tagging requirement results in an efficiency of 70\% for jets containing $b$-hadrons, as determined 
in a simulated sample of $t\bar{t}$ events~\cite{ATL-PHYS-PUB-2016-012}. An overlap removal procedure is 
applied in order to resolve ambiguities between reconstructed physics objects. Jets within $\Delta R=0.2$ of a 
reconstructed electron are removed. If the nearest remaining jet is within $\Delta R = 0.4$ of an electron, the 
electron is removed.  Selected muons with an angular separation of $\Delta R<\min(0.4, 0.04+10 ~\text{GeV}/\pt^{\mu})$ 
from the nearest jet are removed if the jet has at least three tracks originating from the primary vertex; otherwise 
the jet is removed and the muon is kept.  The missing transverse momentum, \met, vector is the negative of the 
vector sum of the transverse momenta of all electrons, muons, and jets. Tracks from the primary vertex\footnote{Proton-proton 
collision vertices are reconstructed by requiring that at least two tracks with $\pt > 0.4$~GeV are associated 
with a given vertex. The primary vertex is defined as the vertex with the largest $\sum p^{2}_\text{T,track}$.} 
that are not associated with any objects are also taken into account in the \met reconstruction~\cite{PERF-2016-07}.

\section{Event selection} 
\label{sec:EventSelection}

Events are required to pass single-lepton or dilepton triggers~\cite{TRIG-2016-01} with minimum 
\pt thresholds in the range 20--26~GeV, depending on the data collection period, and to have at 
least two leptons ($e$ or $\mu$). Events are also required to have at least one lepton (two leptons) 
to be matched to the single-lepton (dilepton) trigger signatures. A higher \pt requirement than the 
online trigger \pt threshold is applied to the trigger-matched lepton.  Three channels are defined 
according to the number of reconstructed leptons (two leptons, three leptons and four leptons), and 
events are further classified according to the charge and flavour of the leptons. In order to suppress 
top quark backgrounds and to be orthogonal to other Higgs boson pair production searches 
($bb\gamma\gamma$~\cite{HIGG-2016-15}, $bbbb$~\cite{EXOT-2016-31}, and $bb\tau\tau$~\cite{HIGG-2016-16}) 
at ATLAS, events containing $b$-tagged jets are rejected.  

\begin{figure}[!ht]
  \centering
   \subfloat[\label{fig:vars-presel-2l-lowNjet-mll-emu-evtsec}]{
    \includegraphics[width=0.32\textwidth]{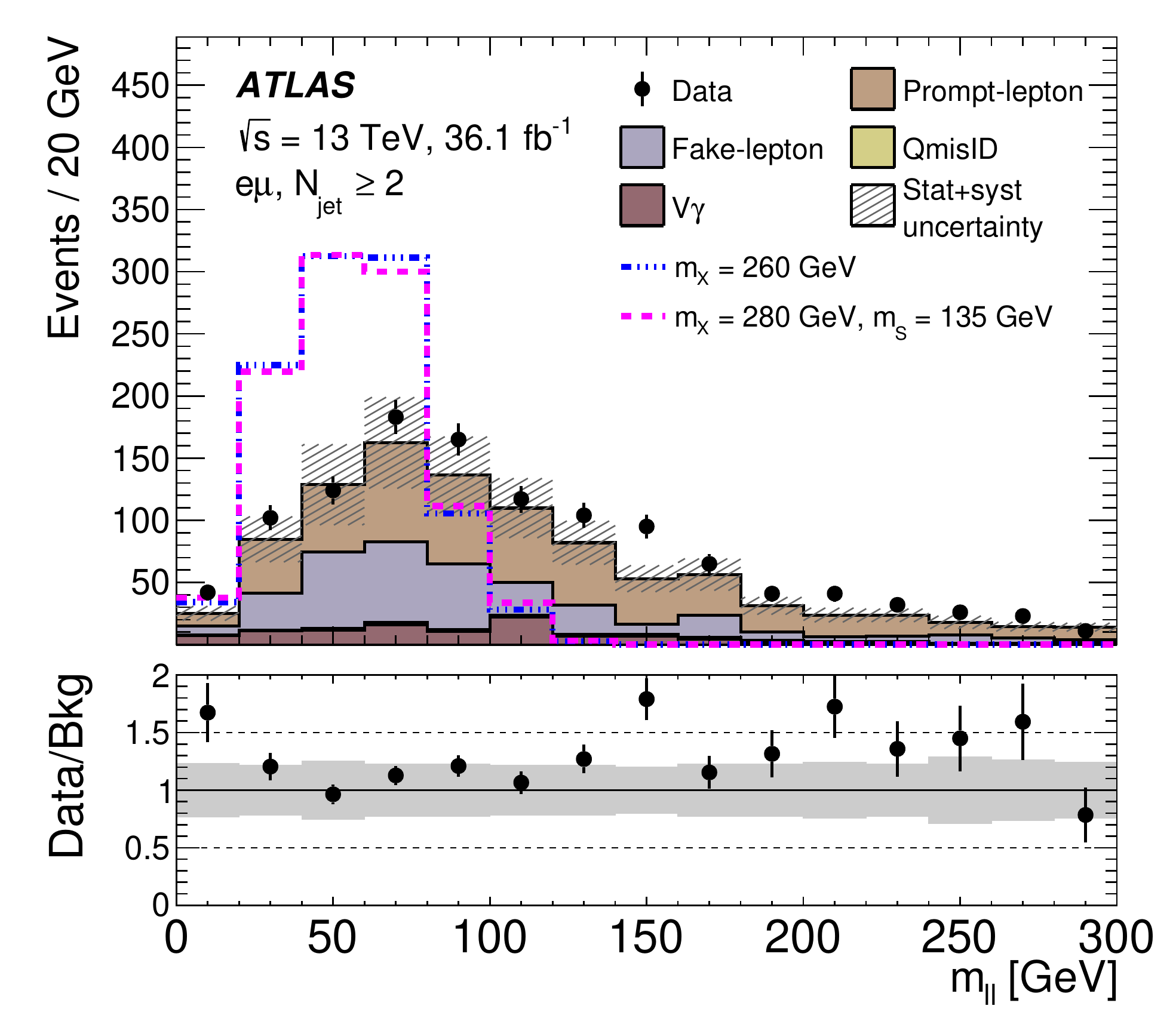}
  }
  \subfloat[\label{fig:vars-presel-3l-mlll-evtsec}]{
    \includegraphics[width=0.32\textwidth]{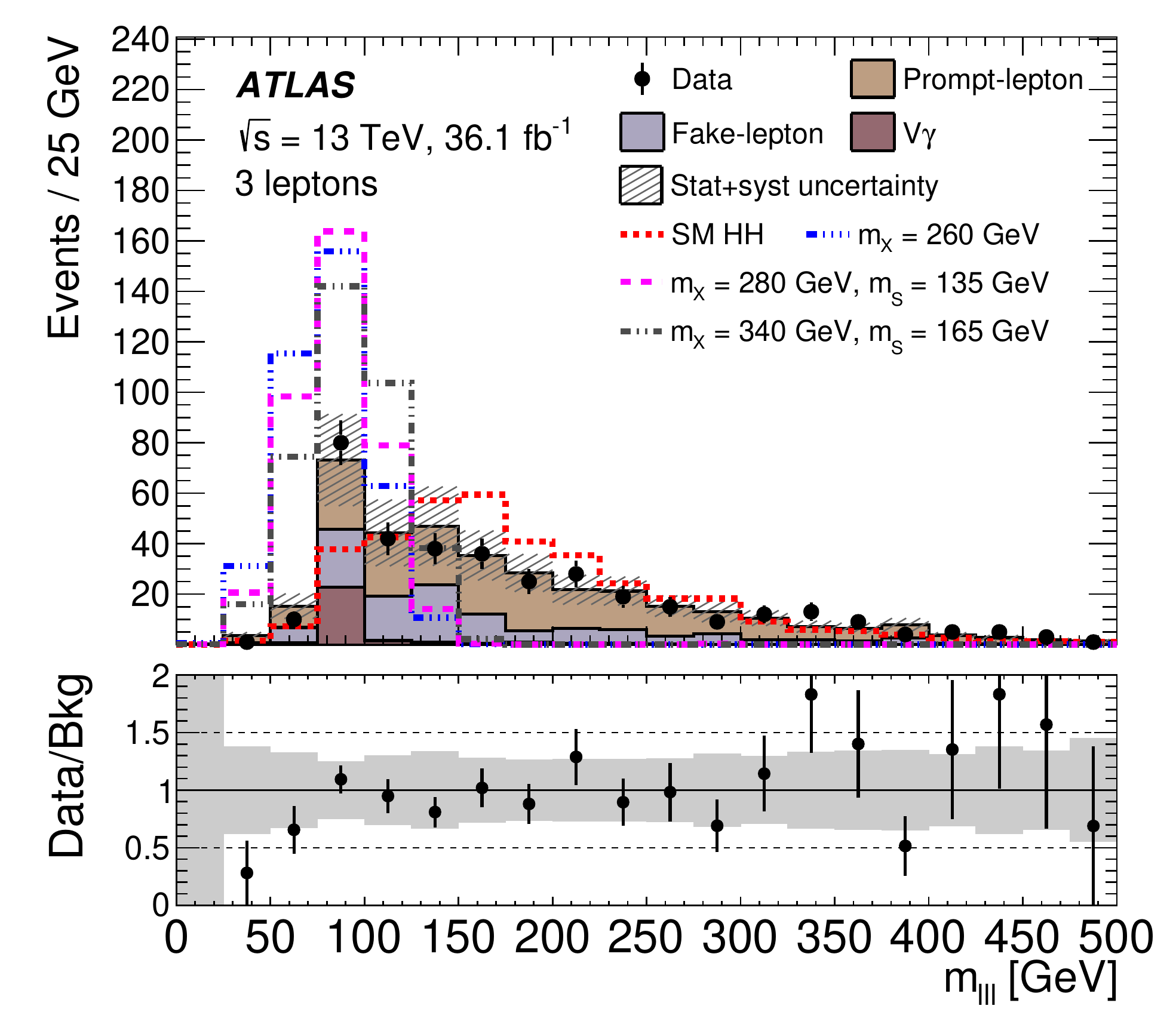}
  }
  \subfloat[\label{fig:vars-presel-4l-2SFOS-m4l-evtsec}]{
    \includegraphics[width=0.32\textwidth]{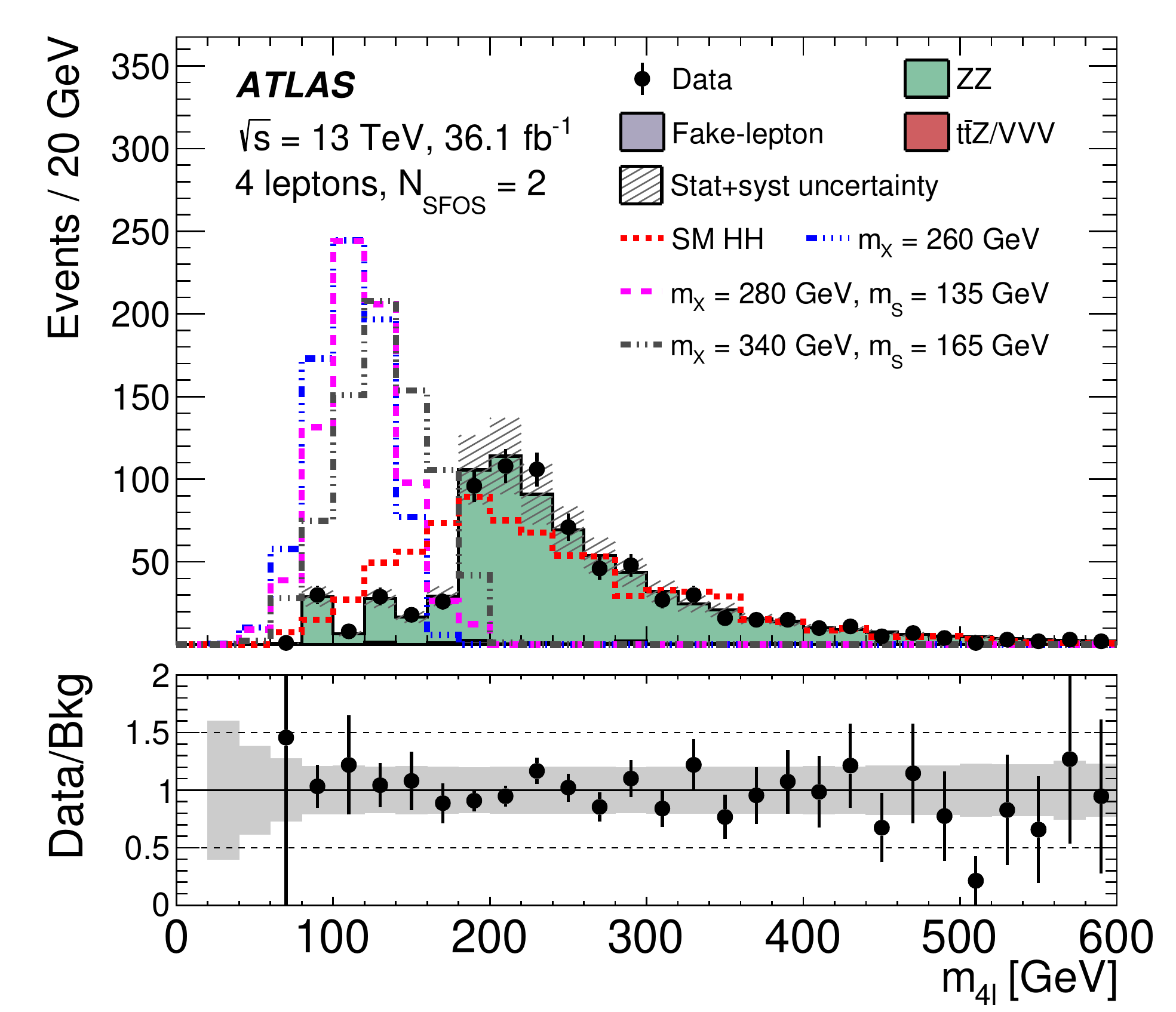}
  } 
  \caption{ 
   Distributions of the invariant mass of \protect\subref{fig:vars-presel-2l-lowNjet-mll-emu-evtsec}~two, 
   \protect\subref{fig:vars-presel-3l-mlll-evtsec}~three, and \protect\subref{fig:vars-presel-4l-2SFOS-m4l-evtsec}~four 
   leptons for the two, three, and four lepton channels after preselection. 
   The charge misidentification background in the two lepton channel and the non-$ZZ$ backgrounds 
   in the four lepton channel are non-zero but are too small to be seen in the distributions.  
   The shaded band in the ratio plot shows the systematic uncertainty in the background estimate. Resonant 
   $HH$ signal samples are denoted by $m_X$.  The integral of each signal sample distribution is scaled to that of the 
   expected background.}
  \label{fig:vars-presel}
\end{figure}

Events in the two lepton channel are required to have exactly two leptons with the same electric charge, 
while the three lepton channel events are required to have exactly three leptons with a summed electric 
charge ${\sum_{i\in\ell} q_i = \pm 1}$. Events are required to have $N_\text{jets} \ge2$ and 
$\met >$~10~(30)~GeV for the two (three) lepton channel. In order to suppress backgrounds containing 
a $Z$ boson in the same-sign $ee$ channel (due to the misidentification of an electron's charge) and in 
the three lepton channel, events are removed if they contained a same-flavour lepton pair with an invariant 
mass, $m_{\ell\ell}$, near the $Z$ boson mass: $|m_{\ell\ell} - m_Z| < 10$~GeV. In order to reduce the 
backgrounds from non-prompt leptons, the leading (subleading) lepton is required to have $\pt>$~30~(20)~GeV 
in the two lepton channel. The two leptons with the same charge are both required to have $\pt>20$~GeV 
in the three lepton channel. For non-resonant production and resonant production with $m_X > 300$~GeV, 
signal events tend to have jets with larger \pt compared to low $m_X$ resonant production scenarios and 
thus $N_\text{jets} \ge3$ is required in the two lepton channel to account for more jets passing the \pt requirement. 
Additionally, events containing a same-flavour opposite-sign (SFOS) lepton pair with an invariant mass 
$m_{\ell\ell} < 15$~GeV are also removed in order to suppress backgrounds from hadron resonances or 
virtual photons. Following this preselection, a number of observables are considered and four variables are chosen 
based on the ranking of the generic algorithm~\cite{Hocker:2007ht} and the correlations betweeen variables. 
These four variables that consist of the angular separation 
between each lepton and the nearest jet as well as invariant masses among different combinations of the 
leptons and jets are used for further selection. The final selections on these variables are optimised 
in order to maximise signal significance. One of these variables is the invariant mass of two (three) leptons in the two (three) lepton 
channel and is shown in Figure \ref{fig:vars-presel-2l-lowNjet-mll-emu-evtsec} (\ref{fig:vars-presel-3l-mlll-evtsec}) 
to illustrate its discriminating power.  The optimisation procedure using all four variables is performed 
separately for each analysis channel, each signal mass point, each lepton flavour category (for the two 
lepton channel), and each number of same-flavour opposite-sign ($N_\text{SFOS}$) lepton pairs (for the 
three lepton channel). The optimised selection criteria are listed in Tables~\ref{tab:vars-2l}--\ref{tab:cuts-3l-XSS} 
in the Appendix. 

Events in the four lepton channel are required to have exactly four leptons with ${\sum_{i\in\ell} q_i = 0}$.  At 
least one of the leptons is required to have $p_\text{T}>$ 22~GeV. Events that contain a SFOS lepton pair with 
$m_{\ell\ell} < 4$~GeV are removed. Following this preselection, selections on the invariant masses and angular 
separation of lepton pairs are implemented to reject backgrounds containing a $Z$ boson or non-prompt leptons 
or other objects incorrectly identified as leptons, known as fake leptons. A summary of the selection criteria used 
in the four lepton channel is shown in Tables~\ref{tab:vars-4l_var_def}--\ref{tab:cuts-4l} in the Appendix. 
Figure~\ref{fig:vars-presel-4l-2SFOS-m4l-evtsec} shows the kinematic distribution of the four lepton invariant mass.

\section{Background estimation} 
\label{sec:BackgroundEstimation}

The backgrounds in this search all have final states that contain leptons that can be classified according to 
their origin into prompt leptons\footnote{Leptons not from hadron decays or photon conversions}, leptons 
with misidentified charges, and fake leptons (including non-prompt and misidentified jets). The backgrounds 
in the two and three lepton channels are dominated by irreducible prompt-lepton processes, including $VV$ 
($WZ$ and $ZZ$), $\ttbar Z$ and $VVV$, with a significant contribution from fake leptons.  The background 
in the four lepton channel is almost exclusively due to $ZZ$ production (including both on-shell and off-shell 
production).

Prompt-lepton backgrounds are modelled using simulated samples described in Section~\ref{sec:DataMC}. 
Control regions containing one pair (two pairs) of SFOS leptons with invariant mass $|m_{\ell\ell} - m_Z| < 10$~GeV 
in the three (four) lepton channel are used to check the modelling of $WZ$ ($ZZ$) background. A data-driven 
method~\cite{HIGG-2018-13,HIGG-2017-03} is used to estimate the charge misidentification rate for electrons 
from a sample of $Z\rightarrow ee$ events with $m_{ee}$ in a narrow window around $m_Z$ . The corresponding 
same-sign charge misidentification (QmisID) background is evaluated by scaling the opposite-sign events by this rate. 
The probability of misidentifying the charge of a muon is checked in both data and simulation, and found to be negligible 
in the kinematic ranges relevant to this analysis.

In the two and three lepton channels non-prompt-lepton contributions from the conversion of prompt photons 
are estimated using $V\gamma$ simulated samples. Fake-lepton and non-prompt-lepton contributions from 
misidentification of hadronic jets as leptons, semileptonic decay of heavy-flavour hadrons and photon conversions 
from neutral pion decays are estimated using data with a fake-factor method~\cite{HIGG-2013-26}. The method 
defines ``tight'' leptons as leptons passing all requirements described in Section~\ref{sec:Objects} and 
``anti-tight'' leptons as leptons failing the isolation or identification requirements. The fake factor is calculated 
as the ratio of events with tight leptons to events with one tight lepton replaced by an anti-tight lepton in the data 
control samples. The control samples of the two and three lepton channels are ensured to be largely orthogonal 
to corresponding preselection samples by requiring a lower jet multiplicity. A control sample containing three leptons 
with enriched $Z$+jets processes is used in the four lepton channel to extract the fake factors. All simulated prompt-lepton 
contributions are subtracted from the data before measuring the fake factor. The fake-lepton background 
contributions are estimated by applying the fake factors to events with the same selection as for the signal regions 
but with at least one anti-tight lepton replacing one of the prompt leptons. The fake factors in the four lepton channel 
are applied to events in two control samples, one with three tight leptons and one anti-tight lepton and the other with 
two tight leptons and two anti-tight leptons.

\section{Systematic uncertainties} 
\label{sec:Systematics}

Experimental systematic uncertainties are evaluated. They include uncertainties related to the electron 
and jet energy measurements~\cite{PERF-2016-04}, muon momentum measurement, \met 
modelling~\cite{PERF-2016-07}, and lepton reconstruction, identification, and isolation efficiencies. The 
dominant systematic uncertainty in the fake-lepton background estimations arises from a closure test of 
the fake-factor method and the relative contributions from heavy-flavour hadron decays and photon 
conversions.  Pileup modelling, $b$-tagging efficiencies, and jet pileup rejection modelling are also included. 
Theoretical uncertainties are evaluated for all simulated samples.  These include uncertainties in PDF, QCD 
scale, and parton shower modelling that impact efficiency times acceptance for signal samples and uncertainties 
in the production cross-sections for simulated background samples. The statistical uncertainties in MC signal 
and background samples as well as in data control regions are included as systematic uncertainties.

The systematic uncertainties with the largest impact on the $HH$ production cross-section (times 
branching ratio) limits come from the jet energy scale and resolution with a relative impact compared 
to the total systematic plus statistical uncertainty of 45\% (29\%--55\%) and fake-lepton background 
estimations with a relative impact of 42\%~(31\%--54\%) for the non-resonant (resonant) production searches.
Theoretical uncertainties are found to have a relative impact of 23\%~(24\%--36\%) for the non-resonant 
(resonant) production searches.  The relative impact of jet energy measurements, fake-lepton background 
estimations, and theoretical uncertainties in the $X\rightarrow SS$ analysis are 38\%--51\%, 37\%--52\% 
and 25\%--32\%, respectively.  Other experimental uncertainties due to lepton, pileup, $b$-tagging, pileup 
jet rejection, prompt-lepton background estimations, and \met modelling are found to have a small impact 
on the results. The uncertainty in the combined 2015+2016 integrated luminosity is 2.1\%. It is derived, 
following a methodology similar to that detailed in Ref.~\cite{DAPR-2013-01}, and using the LUCID-2 detector 
for the baseline luminosity measurements~\cite{LUCID2}, from calibration of the luminosity scale using 
$x$--$y$ beam-separation scans.  It has a 5\%--10\% relative impact due to its simultaneous effect on the 
signal and background estimates.  All simulated processes except $ZZ$ are affected by the uncertainty in 
the luminosity measurement. The relative impact  of all systematic uncertainties is found to be 71\%~(60\%--79\%) 
for the non-resonant (resonant) production searches. In addition to the systematic effects, the statistical 
uncertainties are found to have a relative impact of 71\%~(61\%--80\%) for the non-resonant (resonant) 
production searches.

\section{Results} 
\label{sec:Results} 

\begin{figure}[!htbp]
  \centering
  \includegraphics[width=0.7\textwidth]{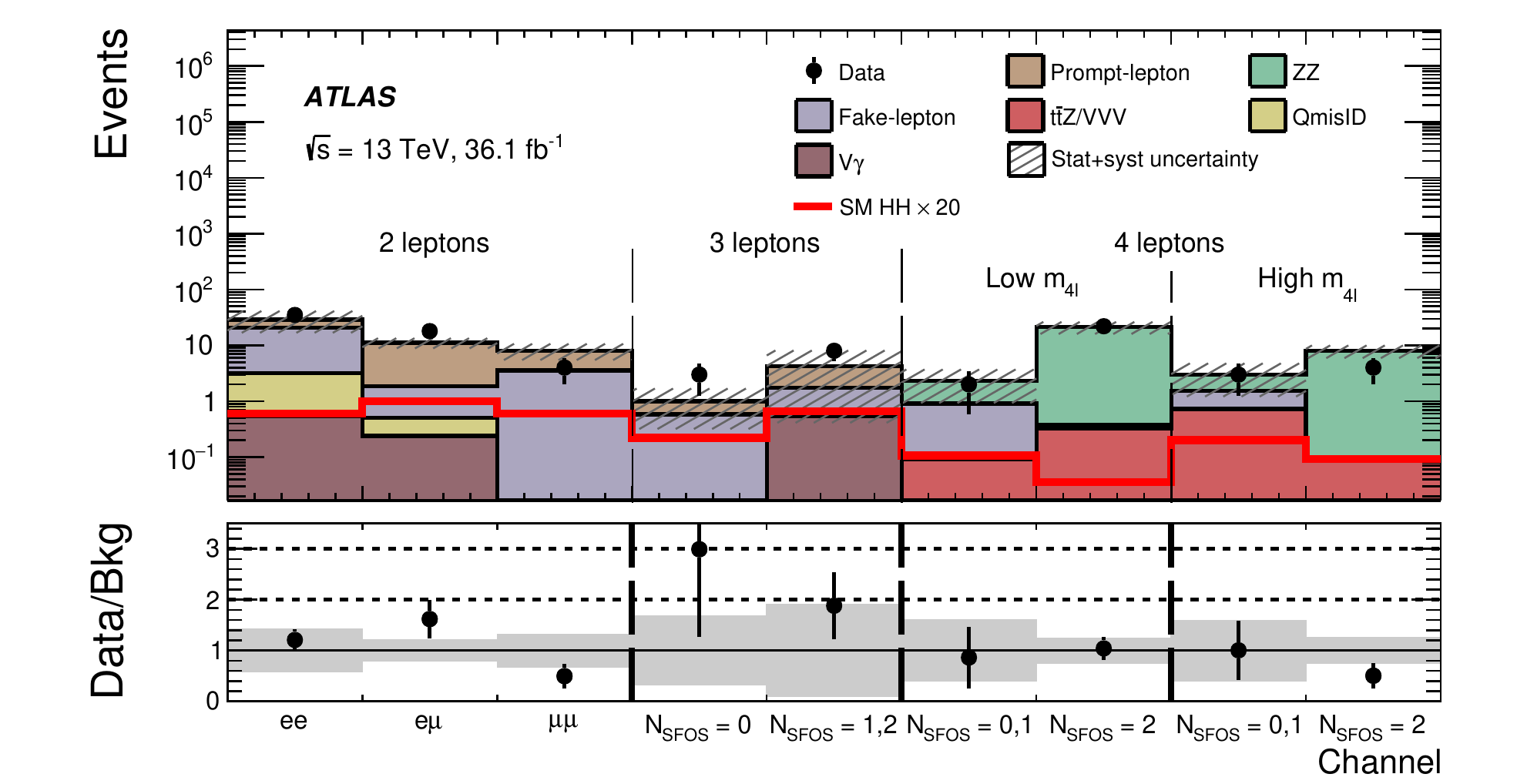}
  \caption{Expected and observed yields in each channel after all selection criteria 
   for the non-resonant $HH$ production searches.  The label $N_\text{SFOS}$ indicates the 
   number of same-flavour, opposite-sign lepton pairs in the channel. Low and high $m_{4\ell}$ 
   indicates $m_{4\ell} < 180$~GeV and $m_{4\ell} > 180$~GeV, respectively. The shaded band 
   in the ratio plot shows the systematic uncertainty in the background estimate. The signal is 
   scaled by a factor of 20. }
  \label{fig:yields-nonres}
\end{figure}

The expected and observed yields in each channel after all selection criteria for the 
non-resonant $HH$ production searches are shown in Figure~\ref{fig:yields-nonres} 
and Table~\ref{tab:yields-nonres_main}.

\begin{table}[!htbp]
  \centering
\begin{tabular}{cc S[table-format=3.2]@{$\pm$}S[table-format=3.2] S[table-format=2.4]@{$\pm$}S[table-format=2.4] S[table-format=1.0]}
\toprule
Channel & Category& \multicolumn{2}{c}{Background} & \multicolumn{2}{c}{Expected Signal} & \multicolumn{1}{c}{Observed} \\
\midrule
    \midrule
 \multirow{3}{*}[-6pt]{2 leptons}
   & $ee$ & 29 & 10 & 0.028 & 0.004 & 35 \\
    \cmidrule{2-7}
    & $e\mu$ & 11.1 & 2.2 & 0.049 & 0.005 & 18 \\
    \cmidrule{2-7}
    & $\mu\mu$ &8.1 & 2.5 & 0.034 & 0.004 & 4 \\
 \midrule
    \midrule
 \multirow{2}{*}[-6pt]{3 leptons}
   & $N_\text{SFOS} = 0$ & 1.0 & 0.7 & 0.011 & 0.005 & 3 \\
    \cmidrule{2-7}
    & $N_\text{SFOS} = \text{1,2}$ & 4.3 & 3.8 & 0.033 & 0.010 & 8 \\
 \midrule
    \midrule

\multirow{2}{*}[-6pt]{4 leptons $m_{4\ell}< 180$~GeV}
   & $N_\text{SFOS} = \text{0,1}$ & 2.3 & 1.4 & 0.005 & 0.001 & 2 \\
    \cmidrule{2-7}
   &$N_\text{SFOS} = 2$ & 21 & 5 & 0.002 & 0.001 & 22 \\

\midrule
    \midrule

\multirow{2}{*}[-6pt]{4 leptons $m_{4\ell}> 180$~GeV}
   & $N_\text{SFOS} = \text{0,1}$ & 3.0 & 1.8 & 0.010 & 0.002 & 3 \\
    \cmidrule{2-7}
    & $N_\text{SFOS} = 2$ & 7.9 & 2.0 & 0.005 & 0.001 & 4 \\

\bottomrule
\end{tabular}
\caption{Expected and observed yields in each channel after all selection criteria and the profile-likelihood fit for the 
              non-resonant $HH$ production searches. The expected signal refers to the SM non-resonant $HH$ production, corresponding 
              to its calculated cross-section at $\sqrt{s}=13$~TeV of 33.4~fb.  The label $N_\text{SFOS}$ indicates 
              the number of same-flavour, opposite-sign lepton pairs in the channel.  Systematic uncertainties on the 
              signal and background estimates are shown.}
  \label{tab:yields-nonres_main}
\end{table}

A statistical analysis using a profile-likelihood-ratio test statistic~\cite{asym} for the two, three, and four 
lepton channels, separately, as well as the combination of the three channels is performed.  The expected 
and observed yields in each of the nine signal regions shown in Figure~\ref{fig:yields-nonres} as well as 
the $ZZ$ control region in the four lepton channel are used as the input parameters to the likelihood. No 
significant excess over the estimated backgrounds is observed in data.  Upper limits at 95\% confidence 
level (CL) are set on the production cross-section for non-resonant SM $HH$ production and on the 
production cross-section times branching ratio for resonant $HH$ production as well as $X\rightarrow SS$ 
production. The expected and observed limits on the signal strength of non-resonant SM $HH$ production, 
defined as the ratio of the signal cross-section to the Standard Model prediction ($\sigma$/$\sigma_\text{SM}$), 
are calculated using the modified frequentist CL$_\text{s}$ method~\cite{cls} using the asymptotic 
approximation and are shown in Table~\ref{tab:limits_nonres}.  All systematic uncertainties are included in 
the profile-likelihood fit as Gaussian nuisance parameters and are treated as correlated across all signal 
regions. The combined observed (expected) upper limit on the non-resonant SM $HH$ production cross-section 
is found to be 5.3~(3.8)~pb, which corresponds to a limit on the signal strength of 160~(120). 

\begin{table}[!htbp]
  \centering
  \begin{tabular}{l S[table-format=3.0] S[table-format=3.0] S[table-format=3.0] S[table-format=3.0] S[table-format=3.0] S[table-format=3.0]}
    \toprule
    & \multicolumn{1}{c}{Observed} & \multicolumn{5}{c}{Expected limit on $\sigma$/$\sigma_\text{SM}$}  \\
    & \multicolumn{1}{c}{limit on $\sigma$/$\sigma_\text{SM}$} & \multicolumn{1}{c}{Median} & \multicolumn{1}{c}{$+2\sigma$} & \multicolumn{1}{c}{$+1\sigma$} & \multicolumn{1}{c}{$-1\sigma$} & \multicolumn{1}{c}{$-2\sigma$} \\
    \midrule
    2 leptons & 170 & 150 & 290 & 210 & 100 & 78 \\
    3 leptons & 420 & 270 & 690 & 420 & 200 & 150 \\
    4 leptons & 340 & 400 & 880 & 590 & 290 & 210 \\
    \midrule
    Combined & 160 & 120 & 230 & 170 & 83 & 62 \\
    \bottomrule
  \end{tabular}
  \caption{Expected and observed 95\% CL exclusion limits set on the non-resonant $HH$
  signal strength. The SM non-resonant $HH$ cross-section at $\sqrt{s}=13$~TeV is calculated to be 33.4~fb.
  Limits are shown for each channel individually as well as for the combination of the channels.  Statistical and systematic uncertainties are included.}
  \label{tab:limits_nonres}
\end{table}

\begin{figure}[!htbp]
  \centering
  \includegraphics[width=0.45\textwidth]{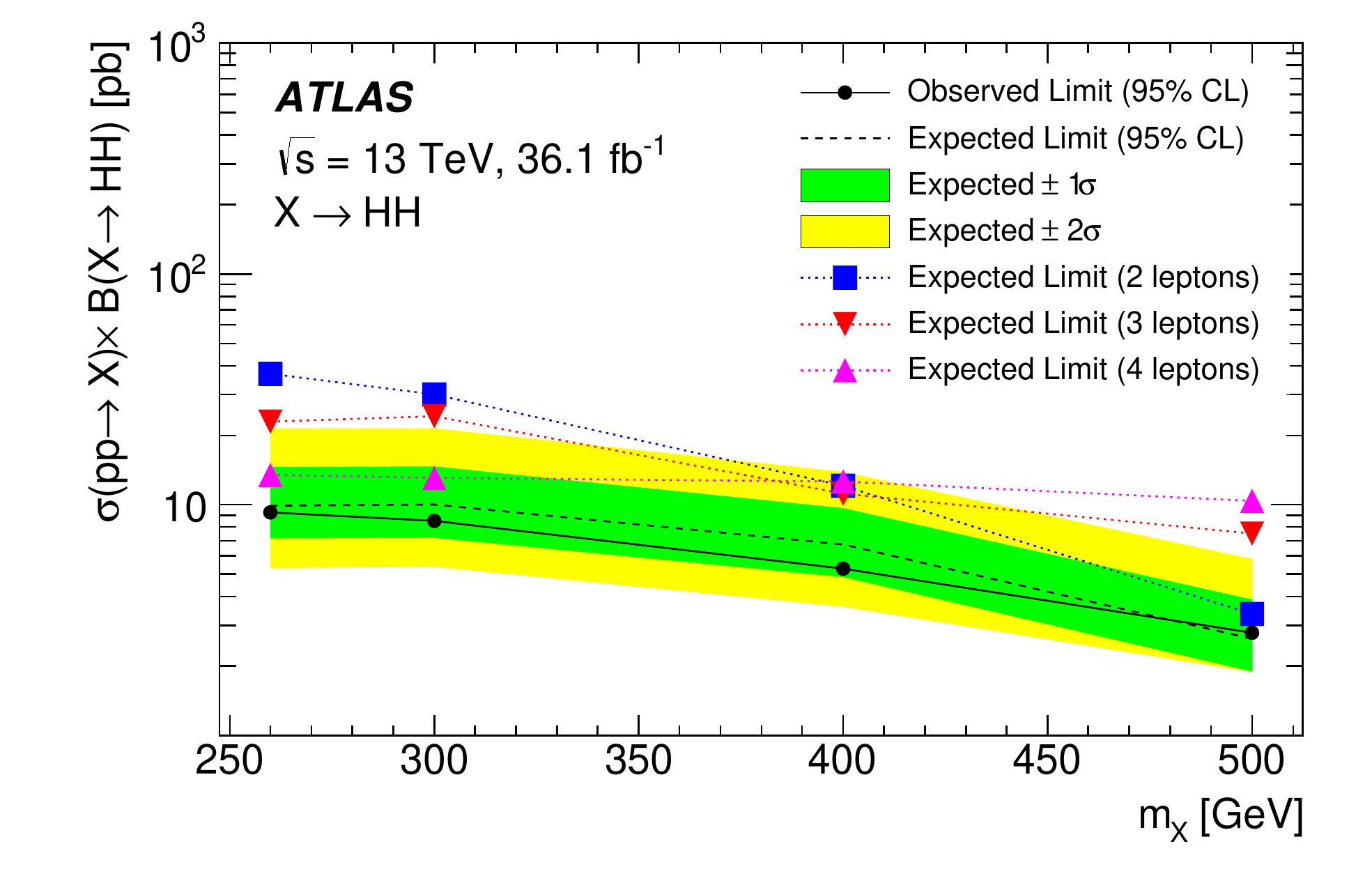}
  \caption{Expected and observed 95\% CL exclusion limits set on the cross-section 
  times branching ratio of resonant $HH$ production as a function of $m_X$.  
  Limits are shown for each channel individually as well as for the combination 
  of the channels.  Statistical and systematic uncertainties are included.}
  \label{fig:limits-res}
\end{figure}

\begin{figure}[!htbp]
  \centering
  \subfloat[\label{fig:limits-XSS-mS}]{
    \includegraphics[width=0.45\textwidth]{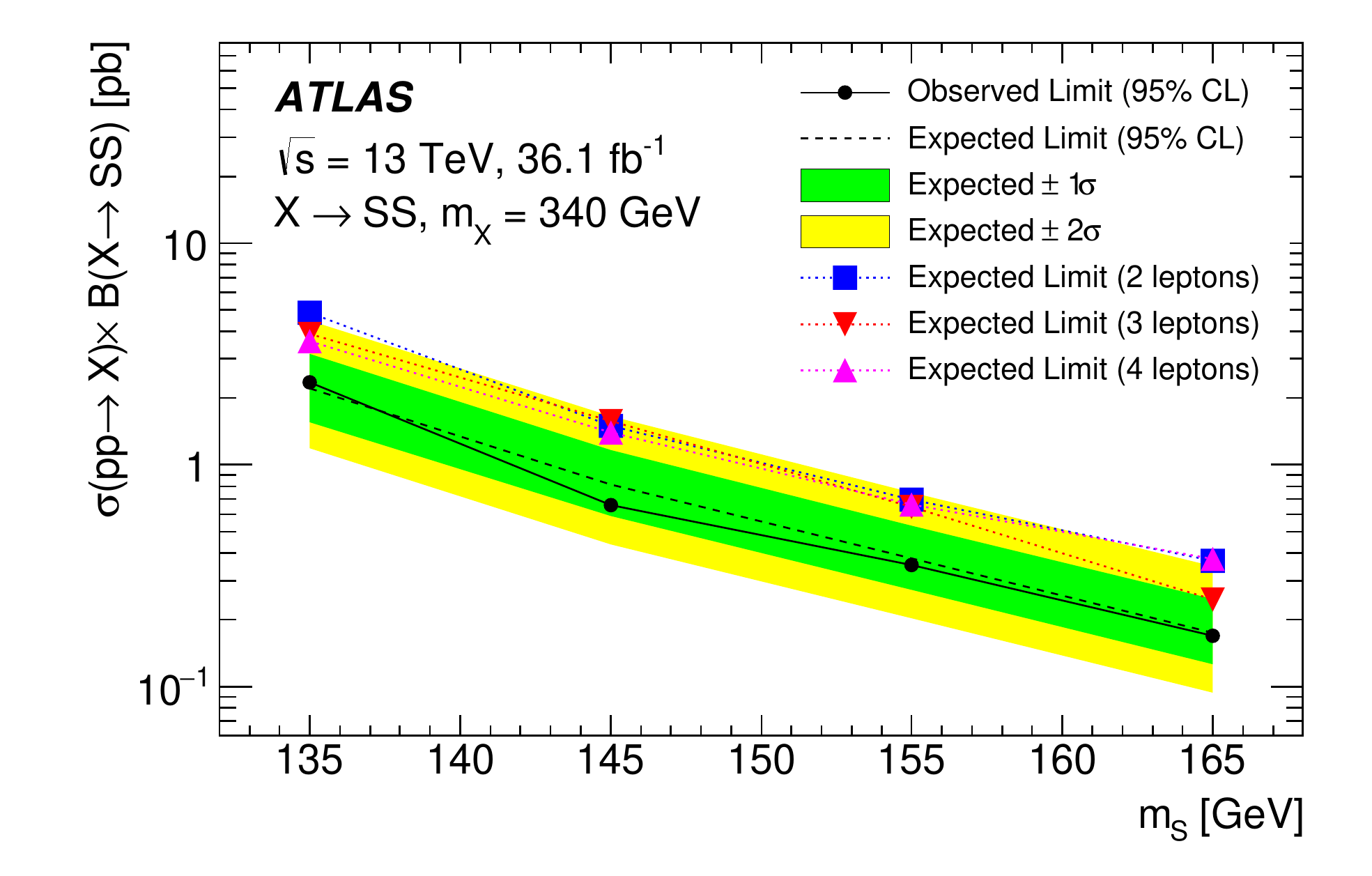}
  }
  \subfloat[\label{fig:limits-XSS-mX}]{
    \includegraphics[width=0.45\textwidth]{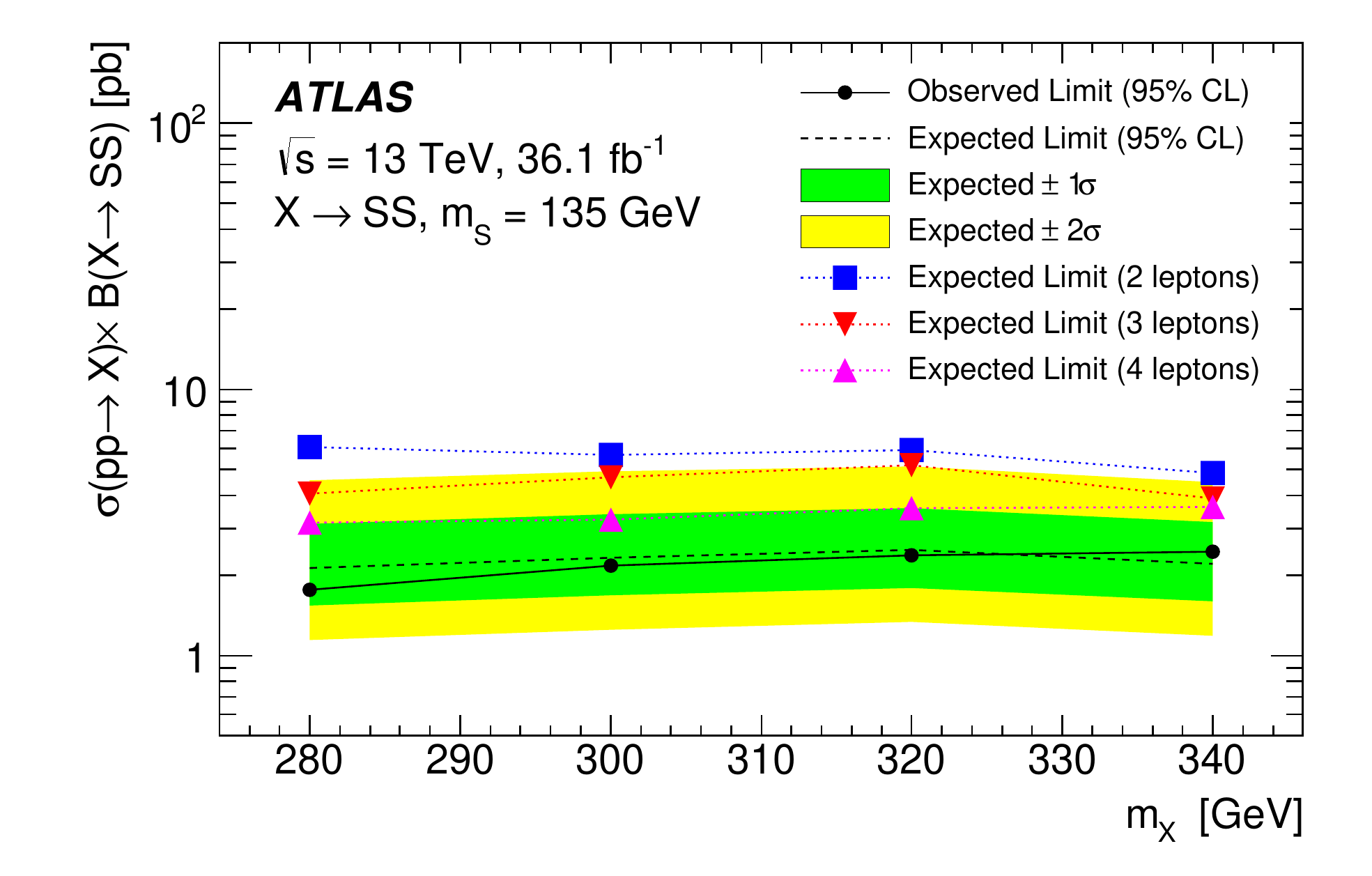}
  }
  \caption{Expected and observed 95\% CL exclusion limits set on the cross-section 
  times branching ratio of  resonant $X\rightarrow SS$ production as a function 
  of~\protect\subref{fig:limits-XSS-mS} $m_S$  and~\protect\subref{fig:limits-XSS-mX} 
  $m_X$.  Limits are shown for each channel individually as well as for the combination 
  of the channels.  Statistical and systematic uncertainties are included.}
  \label{fig:limits-XSS}
\end{figure}

Upper limits at 95\% CL on the production cross-section times branching ratio are set for a scalar 
resonance decaying into either a pair of SM Higgs bosons (shown in Figure~\ref{fig:limits-res}) or 
into a pair of heavy scalars (shown in Figure~\ref{fig:limits-XSS}). The observed (expected) upper 
limits on resonant $HH$ production vary with the resonance mass $m_X$ and range from 
9.3~(10)~pb to 2.8~(2.6)~pb, with the smallest limit set for $m_X = 500$~GeV. Upper limits on 
resonant $SS$ production vary with the resonance mass $m_X$ and the scalar mass $m_S$. The 
observed (expected) limits range from 2.5~(2.5)~pb to 0.16~(0.17)~pb, with the smallest limit set 
for $m_X = 340$~GeV and $m_S = 165$~GeV.  
\FloatBarrier

\section{Conclusions} 
\label{sec:Conclusion}

A search for resonant and non-resonant Higgs boson pair production as well as for a heavy scalar 
pair production has been performed in the $WW^{(*)}WW^{(*)}$ decay channel using 36.1~$\text{fb}^{-1}$ 
of $\sqrt{s} = 13$~TeV proton-proton collision data collected by the ATLAS experiment at the 
LHC in 2015 and 2016. The analysis is performed separately in three channels based on the number 
of leptons in the final state: two same-sign leptons, three leptons, and four leptons. No significant 
excesses over the expected backgrounds are observed in data and the results from the three 
channels are statistically combined.  An observed (expected) 95\% CL upper limit of 160~(120) 
is set on the signal strength for the non-resonant Higgs boson pair production.  Upper limits are 
set on the production cross-section times branching ratio of a heavy scalar $X$ that decays into 
two Higgs bosons for a mass range of 260~GeV~$\le m_X \le$~500~GeV and the observed 
(expected) limits range from 9.3~(10)~pb to 2.8~(2.6)~pb. Upper limits are also set on the 
production cross-section times branching ratio of a heavy scalar $X$ that decays into two heavy 
scalars $S$ for mass ranges of 280~GeV~$\le m_X \le$~340~GeV and 135~GeV~$\le m_S \le$~165~GeV 
and the observed (expected) limits range from 2.5~(2.5)~pb to 0.16~(0.17)~pb.

\section*{Acknowledgements}
% Acknowledgements for papers with collision data
% Version 24-Oct-2018

% Standard acknowledgements start here
%----------------------------------------------
We thank CERN for the very successful operation of the LHC, as well as the
support staff from our institutions without whom ATLAS could not be
operated efficiently.

We acknowledge the support of ANPCyT, Argentina; YerPhI, Armenia; ARC, Australia; BMWFW and FWF, Austria; ANAS, Azerbaijan; SSTC, Belarus; CNPq and FAPESP, Brazil; NSERC, NRC and CFI, Canada; CERN; CONICYT, Chile; CAS, MOST and NSFC, China; COLCIENCIAS, Colombia; MSMT CR, MPO CR and VSC CR, Czech Republic; DNRF and DNSRC, Denmark; IN2P3-CNRS, CEA-DRF/IRFU, France; SRNSFG, Georgia; BMBF, HGF, and MPG, Germany; GSRT, Greece; RGC, Hong Kong SAR, China; ISF and Benoziyo Center, Israel; INFN, Italy; MEXT and JSPS, Japan; CNRST, Morocco; NWO, Netherlands; RCN, Norway; MNiSW and NCN, Poland; FCT, Portugal; MNE/IFA, Romania; MES of Russia and NRC KI, Russian Federation; JINR; MESTD, Serbia; MSSR, Slovakia; ARRS and MIZ\v{S}, Slovenia; DST/NRF, South Africa; MINECO, Spain; SRC and Wallenberg Foundation, Sweden; SERI, SNSF and Cantons of Bern and Geneva, Switzerland; MOST, Taiwan; TAEK, Turkey; STFC, United Kingdom; DOE and NSF, United States of America. In addition, individual groups and members have received support from BCKDF, CANARIE, CRC and Compute Canada, Canada; COST, ERC, ERDF, Horizon 2020, and Marie Sk{\l}odowska-Curie Actions, European Union; Investissements d' Avenir Labex and Idex, ANR, France; DFG and AvH Foundation, Germany; Herakleitos, Thales and Aristeia programmes co-financed by EU-ESF and the Greek NSRF, Greece; BSF-NSF and GIF, Israel; CERCA Programme Generalitat de Catalunya, Spain; The Royal Society and Leverhulme Trust, United Kingdom. 

The crucial computing support from all WLCG partners is acknowledged gratefully, in particular from CERN, the ATLAS Tier-1 facilities at TRIUMF (Canada), NDGF (Denmark, Norway, Sweden), CC-IN2P3 (France), KIT/GridKA (Germany), INFN-CNAF (Italy), NL-T1 (Netherlands), PIC (Spain), ASGC (Taiwan), RAL (UK) and BNL (USA), the Tier-2 facilities worldwide and large non-WLCG resource providers. Major contributors of computing resources are listed in Ref.~\cite{ATL-GEN-PUB-2016-002}.
%----------------------------------------------

\appendix
\part*{Appendix}
\addcontentsline{toc}{part}{Appendix}

\section*{Signal selection criteria after preselection}
\label{app:exact_cuts}
Tables~\ref{tab:vars-2l}--\ref{tab:cuts-2l-HSS} list the final selection criteria in the two lepton channel. 
Tables~\ref{tab:vars-3l}--\ref{tab:cuts-3l-XSS} present the final selection criteria in the three lepton channel. 
Table~\ref{tab:vars-4l_var_def} defines the variables and Table~\ref{tab:cuts-4l} lists the selection criteria in the four lepton channel.  

The lepton pairing strategy in the four leptons channel is designed to identify the decay of a $Z$ boson in 
order to efficiently reject the dominant $ZZ$ background in events with at least one SFOS lepton pair.  Events 
are classified based on the number of SFOS lepton pairs they contain in order to account for the different 
background composition in each signal region.  

Table~\ref{tab:sig_eff_234} shows the final acceptance and selection efficiencies for the signal samples.

\begin{table}[!ht]
  \centering
  \begin{tabular}{cl}
    \toprule
    Variable & \multicolumn{1}{c}{Description} \\
    \midrule
    $\ell_1$ & Leading lepton \\
    \midrule
    $\ell_2$ & Sub-leading lepton \\
    \midrule
    $\Delta R_{\ell_{N}j}$ & Angular distance between $\ell_N$ and the nearest jet \\
    \midrule
    $m_{\ell\ell}$ & Invariant mass of the two leptons \\
    \midrule
    $m_{\ell_Njj}$ & Invariant mass of $\ell_N$ and the two nearest jets \\
    \midrule
    $m_\text{all}$ & Invariant mass of all objects that pass the selection criteria \\
    \midrule
    \bottomrule
  \end{tabular}
  \caption{Description of the notation used in the two lepton analysis.}
  \label{tab:vars-2l}
\end{table}

\begin{table}[!ht]
  \centering
  \begin{tabular}{cccccc}
    \toprule
    $m_X$ & Channel & $\Delta R_{\ell_{1}j}$ & $m_{\ell\ell}$ [GeV] & $m_{\ell_{1}jj}$ [GeV] & $m_\text{all}$ [GeV] \\
    \midrule
    \midrule
    \multirow{3}{*}[-5pt]{260~GeV}
    & $ee$ & [0.35, 1.85] & $< 100$ & $< 145$ & $< 1100$ \\
    \cmidrule{2-6}
    & $e\mu$ & [0.25, 1.80] & $< 85$ & $< 135$ & $< 650$ \\
    \cmidrule{2-6}
    & $\mu\mu$ & [0.25, 2.10] & $< 80$ & $< 115$ & $< 700$ \\
    \midrule
    \midrule
    \multirow{3}{*}[-5pt]{300~GeV} 
    & $ee$ & [0.35, 1.75] & $< 120$ & $< 160$ & $< 1400$ \\
    \cmidrule{2-6}
    & $e\mu$ & [0.20, 1.80] & $< 135$ & $< 160$ & $< 800$ \\
    \cmidrule{2-6}
    & $\mu\mu$ & [0.20, 1.75] & $< 115$ & $< 185$ & $< 1000$ \\
    \bottomrule
  \end{tabular}
  \caption{Optimised selection criteria used in the two lepton channel in the $X\rightarrow HH$ search with $m_X = 260$~GeV and $m_X = 300$~GeV.}
  \label{tab:cuts-2l-lowmass}
\end{table}

\begin{table}[!ht]
  \centering
  \begin{tabular}{cccccc}
    \toprule
    $m_X$ & Channel & $\Delta R_{\ell_{2}j}$ & $\Delta R_{\ell_{1}j}$ & $m_{\ell\ell}$ [GeV] & $m_{\ell_{1}jj}$ [GeV] \\
    \midrule
    \midrule
    \multirow{3}{*}[-5pt]{400~GeV}
    & $ee$ & [0.35, 1.50] & [0.30, 1.25] & [45, 235] & [40, 285] \\
    \cmidrule{2-6}
    & $e\mu$ & [0.20, 1.50] & [0.20, 1.05] & [35, 195] & [30, 235] \\
    \cmidrule{2-6}
    & $\mu\mu$ & [0.20, 1.20] & [0.20, 1.20] & [40, 215] & [30, 260] \\
    \midrule
    \midrule
    \multirow{3}{*}[-5pt]{500~GeV}
    & $ee$ & [0.20, 1.15] & [0.20, 1.15] & [100, 270] & [40, 285] \\
    \cmidrule{2-6}
    & $e\mu$ & [0.20, 1.00] & [0.20, 0.80] & [75, 250] & [35, 350] \\
    \cmidrule{2-6}
    & $\mu\mu$ & [0.20, 1.05] & [0.20, 0.75] & [60, 250] & [30, 310] \\
    \midrule
    \midrule
    \multirow{3}{*}[-5pt]{Non-res.}
    & $ee$ & [0.20, 1.40] & [0.20, 1.15] & [55, 270] & [40, 285] \\
    \cmidrule{2-6}
    & $e\mu$ & [0.20, 1.15] & [0.20, 0.80] & [75, 250] & [35, 350] \\
    \cmidrule{2-6}
    & $\mu\mu$ & [0.20, 1.05] & [0.20, 0.75] & [60, 250] & [30, 310] \\
    \bottomrule
  \end{tabular}
  \caption{Optimised selection criteria used in the two lepton channel in the non-resonant $HH$ search and the $X\rightarrow HH$ search with $m_X = 400$~GeV and $m_X = 500$~GeV.}
  \label{tab:cuts-2l-highmass}
\end{table}

\begin{table}[!ht]
  \centering
  \begin{tabular}{cccccc}
    \toprule
    Mass & Channel & $\Delta R_{\ell_{2}j}$ & $\Delta R_{\ell_{1}j}$ & $m_{\ell\ell}$ [GeV] & $m_{\ell_{1}jj}$ [GeV] \\
    \midrule
    \midrule
    \multirow{3}{*}[-5pt]{$m_S = 135$~GeV}
    & $ee$ & [0.35, 2.5] & [0.4, 1.65] & $< 80$ & [50, 150] \\
    \cmidrule{2-6}
    & $e\mu$ & [0.25, 1.7] & [0.25, 1.65] & $< 95$ & [50, 150] \\
    \cmidrule{2-6}
    & $\mu\mu$ & [0.25, 2.05] & [0.2, 1.85] & $< 95$ & [50, 150] \\
    \midrule
    \midrule
    \multirow{3}{*}[-5pt]{$m_X = 340$~GeV}
    & $ee$ & [0.35, 1.85] & [0.2, 1.65] & $< 130$ & [50, 190] \\
    \cmidrule{2-6}
    & $e\mu$ & [0.25, 1.6] & [0.25, 1.6] & $< 150$ & [50, 150] \\
    \cmidrule{2-6}
    & $\mu\mu$ & [0.2, 2.0] & [0.2, 1.65] & $< 115$ & [50, 185] \\
    \bottomrule
  \end{tabular}
  \caption{Optimised selection criteria used in the two lepton channel in the $X\rightarrow SS$ search.  The selection criteria in the first row are used for $m_S=135$~GeV and $m_X =$~280, 300, and 320~GeV.  The selection criteria in the second row are used for $m_X = 340$~GeV and $m_S =$~135, 145, 155, and 165~GeV.}
  \label{tab:cuts-2l-HSS}
\end{table}

\begin{table}[!ht]
  \centering
  \begin{tabular}{cl}
    \toprule
    Variable & \multicolumn{1}{c}{Description} \\
    \midrule
    $N_\text{SFOS}$ & Number of same-flavour opposite-sign lepton pairs \\
    \midrule
    $\ell_1$ & Lepton with charge opposite to that of the same-sign pair \\
    \midrule
    $\ell_2$ & Lepton from the same-sign pair that is closest to $\ell_1$ in $\eta$--$\phi$ space \\
    \midrule
    $\ell_3$ & Remaining lepton \\
    \midrule
    $m_{\ell\ell\ell}$ & Invariant mass of the three leptons \\
    \midrule
    $m_{\ell_Nj}$ & Invariant mass of $\ell_N$ and the nearest jet \\
    \midrule
    $m_{\ell_Njj}$ & Invariant mass of $\ell_N$ and the two nearest jets \\
    \midrule
    \multirow{2}{*}{$m_{\ell\ell+\ell jj}$} & The minimum sum of the invariant mass of two opposite-sign leptons \\ 
    & and the invariant mass of the remaining lepton and the two leading jets \\
    \midrule
    $\Delta R_{\ell\ell}$ & Angular distance between two leptons \\
    \midrule
    \bottomrule
  \end{tabular}
  \caption{Description of the notation used in the three lepton analysis.}
  \label{tab:vars-3l}
\end{table}

\begin{table}[!ht]
  \centering
  \begin{tabular}{crcc}
    \toprule
    $m_X$ & \multicolumn{1}{c}{Variable} & $N_\text{SFOS} = 0$ & $N_\text{SFOS} = \text{1,2}$ \\
    \midrule
    \midrule
    \multirow{4}{*}[-6pt]{Non-res.}
    & \multicolumn{1}{c}{$\Delta R_{\ell_2\ell_3}$} & [2.47, 5.85] & [2.16, 3.50] \\
    \cmidrule{2-4}
    & $m_{\ell_2\ell_3}$ [GeV] & [10, 70] & [10, 70] \\
    \cmidrule{2-4}
    & $m_{\ell_3jj}$ [GeV] & [50, 110] & [50, 115] \\
    \cmidrule{2-4}
    & $m_{\ell_3j}$ [GeV] & [15, 50] & [15, 45] \\
    \midrule
    \midrule
    \multirow{4}{*}[-6pt]{260}
    & $m_{\ell\ell\ell}$ [GeV] & [30, 105] & [20, 85] \\
    \cmidrule{2-4}
    & $m_{\ell\ell+\ell jj}$ [GeV] & [65, 200] & [85, 360] \\
    \cmidrule{2-4}
    & $m_{\ell_2j}$ [GeV] & [20, 75] & [10, 60] \\
    \cmidrule{2-4}
    & \multicolumn{1}{c}{$\Delta R_{\ell_1\ell_2}$} & [0.58, 1.66] & [0.41, 1.77] \\
    \midrule
    \midrule
    \multirow{4}{*}[-6pt]{300}
    & $m_{\ell\ell\ell}$ [GeV] & [20, 110] & [20, 130] \\
    \cmidrule{2-4}
    & $m_{\ell\ell+\ell jj}$ [GeV] & [55, 195] & [75, 175] \\
    \cmidrule{2-4}
    & $m_{\ell_2j}$ [GeV] & [35, 70] & [15, 85] \\
    \cmidrule{2-4}
    & \multicolumn{1}{c}{$\Delta R_{\ell_1\ell_2}$} & [0.08, 1.49] & [0.42, 1.14] \\
    \midrule
    \midrule
    \multirow{4}{*}[-6pt]{400}
    & $m_{\ell_1\ell_2}$ [GeV] & [20, 60] & [15, 45] \\
    \cmidrule{2-4}
    & $m_{\ell_3j}$ [GeV] & [15, 50] & [15, 50] \\
    \cmidrule{2-4}
    & $m_{\ell\ell+\ell jj}$ [GeV] & [50, 240] & [80, 270] \\
    \cmidrule{2-4}
    & \multicolumn{1}{c}{$\Delta R_{\ell_2\ell_3}$} & [1.97, 6.24] & [2.09, 4.60] \\
    \midrule
    \midrule
    \multirow{4}{*}[-6pt]{500}
    & $m_{\ell\ell\ell}$ [GeV] & [130, 320] & [150, 295] \\
    \cmidrule{2-4}
    & \multicolumn{1}{c}{$\Delta R_{\ell_2\ell_3}$} & [2.68, 3.47] & [2.54, 6.19] \\
    \cmidrule{2-4}
    & \multicolumn{1}{c}{$\Delta R_{\ell_1\ell_2}$} & [0.12, 0.68] & [0.11, 1.08] \\
    \cmidrule{2-4}
    & $m_{\ell_3j}$ [GeV] & [15, 90] & [20, 50] \\
    \bottomrule
  \end{tabular}
  \caption{Optimised selection criteria for non-resonant and resonant $HH$ searches in the three lepton channel.  The selection criteria are chosen to ensure constant signal selection efficiency between the $N_\text{SFOS} = 0$ and $N_\text{SFOS} = \text{1,2}$ categories.}
  \label{tab:cuts-3l-hh}
\end{table}

\begin{table}[!ht]
  \centering
  \begin{tabular}{crcc}
    \toprule
    $m_X$/$m_S$ & \multicolumn{1}{c}{Variable} & $N_\text{SFOS} = 0$ & $N_\text{SFOS} = \text{1,2}$ \\
    \midrule
    \midrule
    \multirow{4}{*}[-8pt]{\shortstack{280  \\ 135}}
    & $m_{\ell\ell\ell}$ [GeV] & [55, 100] & [25, 85] \\
    \cmidrule{2-4}
    & $m_{\ell_3jj}$ [GeV] & [50, 145] & [50, 300] \\
    \cmidrule{2-4}
    & $m_{\ell_2j}$ [GeV] & [35, 75] & [10, 65] \\
    \cmidrule{2-4}
    & \multicolumn{1}{c}{$\Delta R_{\ell_1\ell_2}$} & [0.51, 1.61] & [0.19, 1.16] \\
    \midrule
    \midrule
    \multirow{4}{*}[-8pt]{\shortstack{300  \\ 135}}
    & $m_{\ell\ell\ell}$ [GeV] & [55, 110] & [20, 135] \\
    \cmidrule{2-4}
    & $m_{\ell_3jj}$ [GeV] & [50, 190] & [50, 135] \\
    \cmidrule{2-4}
    & $m_{\ell_2j}$ [GeV] & [20, 55] & [20, 50] \\
    \cmidrule{2-4}
    & \multicolumn{1}{c}{$\Delta R_{\ell_1\ell_2}$} & [0.10, 1.86] & [0.46, 3.38] \\
    \midrule
    \midrule
    \multirow{4}{*}[-8pt]{\shortstack{320 \\ 135}}
    & $m_{\ell\ell\ell}$ [GeV] & [25, 110] & [25, 135] \\
    \cmidrule{2-4}
    & $m_{\ell_3jj}$ [GeV] & [60, 210] & [50, 135] \\
    \cmidrule{2-4}
    & $m_{\ell_2j}$ [GeV] & [10, 55] & [30, 60] \\
    \cmidrule{2-4}
    & \multicolumn{1}{c}{$\Delta R_{\ell_1\ell_2}$} & [0.24, 1.78] & [0.15, 1.53] \\
    \midrule
    \midrule
    \multirow{4}{*}[-8pt]{\shortstack{340 \\ 135}}
    & $m_{\ell\ell\ell}$ [GeV] & [50, 170] & [25, 180] \\
    \cmidrule{2-4}
    & $m_{\ell_3jj}$ [GeV] & [50, 115] & [50, 115] \\
    \cmidrule{2-4}
    & $m_{\ell_2j}$ [GeV] & [10, 40] & [25, 65] \\
    \cmidrule{2-4}
    & \multicolumn{1}{c}{$\Delta R_{\ell_1\ell_2}$} & [0.12, 1.68] & [0.15, 1.10] \\
    \midrule
    \midrule
    \multirow{4}{*}[-8pt]{\shortstack{340 \\ 145}}
    & $m_{\ell\ell\ell}$ [GeV] & [60, 110] & [40, 130] \\
    \cmidrule{2-4}
    & $m_{\ell_3jj}$ [GeV] & [50, 350] & [50, 140] \\
    \cmidrule{2-4}
    & $m_{\ell_2j}$ [GeV] & [10, 55] & [10, 90] \\
    \cmidrule{2-4}
    & \multicolumn{1}{c}{$\Delta R_{\ell_1\ell_2}$} & [0.19, 1.58] & [0.41, 1.11] \\
    \midrule
    \midrule
    \multirow{4}{*}[-8pt]{\shortstack{340 \\ 155}}
    & $m_{\ell\ell\ell}$ [GeV] & [30, 110] & [35, 135] \\
    \cmidrule{2-4}
    & $m_{\ell_3jj}$ [GeV] & [50, 205] & [50, 140] \\
    \cmidrule{2-4}
    & $m_{\ell_2j}$ [GeV] & [20, 55] & [10, 85] \\
    \cmidrule{2-4}
    & \multicolumn{1}{c}{$\Delta R_{\ell_1\ell_2}$} & [0.27, 2.24] & [0.50, 1.15] \\
    \midrule
    \midrule
    \multirow{4}{*}[-8pt]{\shortstack{340 \\ 165}}
    & $m_{\ell\ell\ell}$ [GeV] & [25, 110] & [25, 135] \\
    \cmidrule{2-4}
    & $m_{\ell_3jj}$ [GeV] & [50, 210] & [50, 140] \\
    \cmidrule{2-4}
    & $m_{\ell_2j}$ [GeV] & [15, 55] & [20, 60] \\
    \cmidrule{2-4}
    & \multicolumn{1}{c}{$\Delta R_{\ell_1\ell_2}$} & [0.20, 2.12] & [0.39, 1.95] \\
    \bottomrule
  \end{tabular}
  \caption{Optimised selection criteria for the $X\rightarrow SS$ searches in the three lepton channel.  The selection criteria are chosen to ensure constant signal selection efficiency between the $N_\text{SFOS} = 0$ and $N_\text{SFOS} = \text{1,2}$ categories.}
  \label{tab:cuts-3l-XSS}
\end{table}

\begin{table}[!ht]
  \centering
  \begin{tabular}{cl}
    \toprule
    Variable & \multicolumn{1}{c}{Description} \\
    \midrule
    $\pt^i$& \pt of lepton $i$\\
     \midrule
    $\ell_2$ and $\ell_3$ ($N_\text{SFOS}>0$) & SFOS lepton pair with invariant mass closest to $Z$ boson ($p_\text{T,2}>p_\text{T,3}$) \\ 
    \midrule
    \multirow{2}{*}{$\ell_2$ and $\ell_3$ ($N_\text{SFOS}=0$)} & Different-flavour OS lepton pair with \\
    & invariant mass closest to $Z$ boson ($p_\text{T,2}>p_\text{T,3}$) \\
    \midrule
    $\ell_0$ and $\ell_1$&Remaining lepton pair ($p_\text{T,0}>p_\text{T,1}$) \\
    \midrule
    \bottomrule
  \end{tabular}
  \caption{Description of the notation used in the four lepton analysis.}
  \label{tab:vars-4l_var_def}
\end{table}

\begin{table}[htbp]
  \begin{center}
    \begin{tabular}{c|c}
      \toprule
       \multicolumn{2}{c}{\textbf{Event selection in the four lepton channel} } \\
      \midrule
      \multicolumn{2}{c}{4 leptons with $\pt>$ 10~GeV and $\sum q_i = 0$} \\ 
      \midrule
       \multicolumn{2}{c}{    Trigger} \\
      \midrule
       \multicolumn{2}{c}{Trigger matched lepton} \\ 
      \midrule
       \multicolumn{2}{c}{$\pt^{\ell_\text{matched}} > 22,25,27$~GeV \scriptsize{(depending on data period trigger)} } \\
      \midrule
       \multicolumn{2}{c}{$m_{\ell\ell} > 4$~GeV \scriptsize{(for all SFOS pairs)}} \\ 
      \midrule
       \multicolumn{2}{c}{$N_{b\text{-tag}}=0$} \\
          \midrule
       \multicolumn{2}{c}{$m_{\ell_0\ell_1} >$ 10~GeV} \\
       \midrule
      \midrule
      \multicolumn{2}{c}{ \textbf{$N_\text{SFOS} = \text{0,1}$ selection}} \\
      \midrule
       \multicolumn{2}{c}{$|m_{\ell_2\ell_3} - m_{Z}| > 5$~GeV} \\
         \midrule
        \textbf{$m_{4\ell} < $~180~GeV}&\textbf{$m_{4\ell} > $~180~GeV} \\
      \midrule
      \midrule
    \multicolumn{2}{c}{   \textbf{$N_\text{SFOS} = 2$ selection} } \\
      \midrule
      \multicolumn{2}{c}{ $m_{\ell_2\ell_3} <$~70~GeV, $ m_{\ell_2\ell_3} >$~110~GeV} \\
      \midrule
      \textbf{$m_{4\ell} < $~180~GeV}&\textbf{$m_{4\ell} > $~180~GeV} \\
      \midrule
     ~~~~~~~~~~~~~ $\Delta\phi_{\ell_2\ell_3} < $~2.6~rad ~~~~~~~~~~~~~& $m_{\ell_0\ell_1} <$~70~GeV, $ m_{\ell_0\ell_1} >$~110~GeV \\
      \bottomrule
    \end{tabular}
    \caption{Summary of the selection criteria used in the four lepton channel.  All events are required to pass the common selection and then category-dependent selection criteria are applied according to the number of same-flavour opposite-sign lepton pairs in the event.} 
    \label{tab:cuts-4l}
  \end{center}
\end{table}

\begin{table}[htbp]
\begin{center}
\begin{tabular}{lcccc}
\toprule
       \multirow{3}{*}{Channel}  & \multirow{3}{*}{Category}      & \multirow{3}{*}{Non-resonant $HH$}       & \multirowcell{3}{Resonant $HH$ \\ $m_X \in [280,340]$~GeV}         &$X\rightarrow SS$           \\
       & & & & $m_X \in [280,340]$~GeV           \\
       & & & & $m_S \in [135,165]$~GeV           \\
                &      & [\%]      & [\%]         & [\%]           \\
\midrule
\midrule
\multirow{3}{*}[-5pt]{Two lepton} &$ee$          &0.60   &0.30--0.55   &0.41--0.82       \\
    \cmidrule{2-5}                &$e\mu$        &1.05   &0.52--1.32   &1.12--2.31       \\      
    \cmidrule{2-5}                &$\mu\mu$      &0.66   &0.35--1.10   &0.88--1.94       \\
\midrule
\midrule
\multirow{2}{*}[-5pt]{Three lepton} & $N_\text{SFOS} = 0$  &0.32   & 0.07--0.24    & 0.09--0.5\\
     \cmidrule{2-5}                 & $N_\text{SFOS} = \text{1,2}$  &0.94   & 0.18--0.61   & 0.27--1.2\\
\midrule
\midrule
\multirow{2}{*}[-5pt]{Four lepton} & $N_\text{SFOS} = \text{0,1}$  &  2.94  &  2.08--3.32  & 2.65--3.66 \\
     \cmidrule{2-5}                & $N_\text{SFOS} = 2$          		   &  1.23  &  0.73--1.34  & 0.85--1.46 \\
\bottomrule
\end{tabular}
\caption{The final acceptance times selection efficiencies in the $4W$ channel for non-resonant, resonant, and $SS$ 
              signal samples after all selection criteria are applied.  Acceptance times selection efficiency is defined as the 
              ratio of reconstructed signal events passing all selection criteria to the number of generated signal events that 
              are filtered for the corresponding channel.  The generator filter efficiencies are $4.4\times10^{-3}$ for the two 
              same-sign lepton channel, $4.2\times10^{-3}$ for the three lepton channel, and $5.1\times10^{-4}$ for the four 
              lepton channel.  All numbers are given as percentages.}
\label{tab:sig_eff_234}
\end{center}
\end{table}

\clearpage

\printbibliography

\clearpage 
% ATLAS Collaboration author list
% Reference date of HIGG-2016-24 is 2018-05-17
% Author list last updated on date 14-AUG-18
% Data extracted on 14-Aug-2018 for paper reference HIGG-2016-24
% at 7:22pm
 
\begin{flushleft}
{\Large The ATLAS Collaboration}

\bigskip

M.~Aaboud$^\textrm{\scriptsize 34d}$,    
G.~Aad$^\textrm{\scriptsize 99}$,    
B.~Abbott$^\textrm{\scriptsize 124}$,    
O.~Abdinov$^\textrm{\scriptsize 13,*}$,    
B.~Abeloos$^\textrm{\scriptsize 128}$,    
D.K.~Abhayasinghe$^\textrm{\scriptsize 91}$,    
S.H.~Abidi$^\textrm{\scriptsize 164}$,    
O.S.~AbouZeid$^\textrm{\scriptsize 39}$,    
N.L.~Abraham$^\textrm{\scriptsize 153}$,    
H.~Abramowicz$^\textrm{\scriptsize 158}$
% ...
\bigskip
\\

% ...
$^{13}$Institute of Physics, Azerbaijan Academy of Sciences, Baku; Azerbaijan.\\
% ...
$^{34}$$^{(a)}$Facult\'e des Sciences Ain Chock, R\'eseau Universitaire de Physique des Hautes Energies - Universit\'e Hassan II, Casablanca;$^{(b)}$Centre National de l'Energie des Sciences Techniques Nucleaires (CNESTEN), Rabat;$^{(c)}$Facult\'e des Sciences Semlalia, Universit\'e Cadi Ayyad, LPHEA-Marrakech;$^{(d)}$Facult\'e des Sciences, Universit\'e Mohamed Premier and LPTPM, Oujda;$^{(e)}$Facult\'e des sciences, Universit\'e Mohammed V, Rabat; Morocco.\\
% ...
$^{39}$Niels Bohr Institute, University of Copenhagen, Copenhagen; Denmark.\\
% ...
$^{91}$Department of Physics, Royal Holloway University of London, Egham; United Kingdom.\\
% ...
$^{99}$CPPM, Aix-Marseille Universit\'e, CNRS/IN2P3, Marseille; France.\\
% ...
$^{124}$Homer L. Dodge Department of Physics and Astronomy, University of Oklahoma, Norman OK; United States of America.\\
% ...
$^{128}$LAL, Universit\'e Paris-Sud, CNRS/IN2P3, Universit\'e Paris-Saclay, Orsay; France.\\
% ...
$^{153}$Department of Physics and Astronomy, University of Sussex, Brighton; United Kingdom.\\
% ...
$^{158}$Raymond and Beverly Sackler School of Physics and Astronomy, Tel Aviv University, Tel Aviv; Israel.\\
% ...
$^{164}$Department of Physics, University of Toronto, Toronto ON; Canada.\\
% ...
$^{a}$ Also at Borough of Manhattan Community College, City University of New York, NY; United States of America.\\
$^{b}$ Also at Centre for High Performance Computing, CSIR Campus, Rosebank, Cape Town; South Africa.\\
$^{c}$ Also at CERN, Geneva; Switzerland.\\
$^{d}$ Also at CPPM, Aix-Marseille Universit\'e, CNRS/IN2P3, Marseille; France.\\
$^{e}$ Also at D\'epartement de Physique Nucl\'eaire et Corpusculaire, Universit\'e de Gen\`eve, Gen\`eve; Switzerland.\\
% ...
$^{*}$ Deceased

\end{flushleft}

% Created with Glance <Atlas.Glance@cern.ch>

\end{document}